\documentstyle [12pt] {article}

\def \tablename {\bf Table}
\def\fnum@table{\bf \tablename\ \thetable}

\newcommand{\be}{\begin{equation}}
\newcommand{\ee}{\end{equation}}
\newcommand{\bes}{\begin{eqnarray}}
\newcommand{\ees}{\end{eqnarray}}
\newcommand{\bma}{\left( \begin {array}}
\newcommand{\ema}{\end {array} \right)}
\newcommand{\ra}{\rightarrow}
\def \fhat{\hat{F}}
\def \tr{\rm Tr}

\newcommand{\vx}{\mbox{$\vec{x}$}}
\newcommand{\vy}{\mbox{$\vec{y}$}}
\def \v0{{\vec{0}}}
\def \vx{{\vec{x}}}
\def \vy{{\vec{y}}}
\def \fhat{\hat{f}}

\def \tmin{t_{\rm min}}
\def \chidof{\chi^2/{\rm d.o.f.}}

\def \KroMac{KMc}

\begin{document}

\begin{flushright}
DESY 93-179\\
PSI-PR-93-23\\
WUB 93-39 \\
\end{flushright}

\begin{center}
{\bf
\large
The Leptonic Decay Constants of $\bar{Q}q$ Mesons
and the Lattice Resolution}\footnote{work
supported in part by DFG grant Schi 257/3-1.} \\
\end{center}
\normalsize
\medskip

\begin{center}
C. Alexandrou$^{a}$\footnote {Present address: Department of
Natural Sciences, University of
Cyprus, Nicosia, Cyprus.},
S. G\"usken$^{b}$,  F. Jegerlehner$^c$,
K. Schilling$^b$, G. Siegert$^b$,
R. Sommer$^d$
\\
\end{center}

$^a$ {\it Paul Scherrer Institut, CH-5232 Villigen PSI, Switzerland
     {\rm and} Department of Natural Sciences, University of
Cyprus, Nicosia, Cyprus}\\
$^b$ {\it Physics Department, University of Wuppertal, D-42097
Wuppertal,
          Germany}\\
$^c$ {\it Paul Scherrer Institut, CH-5232 Villigen PSI, Switzerland}\\
$^d$ {\it DESY, Theory Division, D-22603 Hamburg, Germany}\\
\\
\centerline{{\bf Abstract}}
We present a high statistics study
of the leptonic decay constant $f_P$ of heavy pseudoscalar mesons
using
propagating heavy Wilson quarks within the quenched
approximation, on
lattices  covering sizes from about 0.7~fm to 2~fm.
Varying $\beta$ between 5.74 and 6.26
we observe
a sizeable
$a$ dependence of $f_P$ when one uses the quark
field
normalization that was suggested by Kronfeld and Mackenzie,
compared with the weaker
dependence observed for the standard relativistic norm. The two
schemes come into agreement when one extrapolates to $a \rightarrow 0$. The
extrapolations needed to reach the continuum quantity $f_B$
introduce large errors and lead
 to the value $f_B=0.18(5)$~GeV in
the quenched approximation.
This suggests that much more effort will
be needed
to obtain an accurate lattice prediction for $f_B$.

\newpage

\section{Introduction}
The weak decays of $D$- and $B$-mesons will allow to extract the
Cabibbo-Kobayashi-Maskawa matrix elements from the experimental
data, once
the hadronic interactions in the relevant weak matrix elements
are
determined from QCD.
Lattice calculations in principle enable us  to compute these
QCD effects without  model assumptions.

The simulation
of heavy-light quark systems in lattice QCD
requires lattice spacings $a$  smaller than the inverse
relevant masses.
At present one reaches values
$a^{-1}\,\sim$ 2 to 4 GeV such that $D$-meson properties are
amenable to lattice techniques.

The $B$-meson cannot be handled in such a direct way
with lattice methods. The static approximation on the
lattice~\cite{Eichten} enables us to calculate the properties
of pseudoscalar mesons in the limit of infinite quark mass, and
it is
very natural to estimate the $B$-meson matrix elements by
interpolating
between the $D$-mass range and the static point.
The latter has been investigated by various
groups~\cite{Allt1,fb1,fb3,Eichtenstat,BLSnew,UKQCD,Ape},
and it has been shown that
(depending on the way one sets the
scale)
within the range $5.7 < \beta < 6.3$,
the value of the leptonic decay constant $f_{stat}$
changes  significantly with  the lattice
resolution~\cite{fb3,eichtendal}.
Although the conventional {\it i.e.~nonstatic}
treatment of `heavy' quarks (applied within --- and slightly
beyond ---
the charm region) has been pursued by a number of
groups~\cite{conventional},  the  $a$ dependence
of $f_{P}$ has not yet been  systematically studied.

As actual lattice calculations with relativistic heavy quarks
do not respect the inequality  $a m_{q} \ll 1$, lattice artifacts
are an important issue.
It has been suggested that lattice artifacts might be reduced
by a modification of the relativistic normalization of states,
$\sqrt{2 \kappa}$, at order
$O(am_{q})$~\cite{BLSnew,axial,altbernard,KM,NRQCD}.
In particular,
Kronfeld and Mackenzie~\cite{KM} have given
arguments, that
heavy quark fields in the Wilson formulation should be normalized
by
$\sqrt{1-3\kappa/4\kappa_c}$, which differs significantly
from the
naive normalization. This has been shown to yield a much smoother
$1/M_P$ behaviour towards the static point at a fixed
lattice spacing~\cite{BLSnew,UKQCD}.

The real check of improvement
is to observe  a flatter $a$-dependence of $f_P$ at fixed
$m_q$.
In this paper we  address this issue
by presenting  the results of a high statistics  study of
lattices  at
$\beta$= 5.74,~6.00 and 6.26 (with lattice extensions between 0.7
fm to 2 fm and
inverse lattice spacings between 1.2 and 3.2 GeV). The results
were
obtained using standard Wilson fermions in the quenched
approximation. Smearing
techniques were needed to improve the saturation of the weak
current
correlators by the lowest lying states at large times.

The paper is organized as follows: In Section 2 we discuss some
details of the calculation, in particular the smearing of
operators
applied to improve the signals at large time
separations. We present  a   factorization test
among local-local, local-smeared and smeared-smeared correlators
to ensure
ground state dominance.
In Section 3 we briefly describe the Kronfeld-Mackenzie proposal
to normalize heavy-light states.
 Our main  results are
contained in Section 4 where the $a$-dependence
and finite size effects
of the leptonic
decay
constant is presented and the implication of the
Kronfeld-Mackenzie
normalization at finite values of $a$ is discussed.  After
extrapolation to the continuum, we perform the interpolation
between
conventional results and the static point, to obtain the estimate
for
$f_B$.  Summary and conclusions follow in Section 5.


\section{Operators and Smearing}

The starting point of this paper is our
recent  study of
leptonic decay constants of heavy-light mesons in the static
quark
limit~\cite{fb3}. There, we optimized our trial wave functions
with respect to maximal ground state dominance.
In the following we will use the  `best' wave function, i.e. a
`gaussian type'
wave function with parameters $n=100$ and $\alpha=4$, from
ref.~\cite{fb3}, where the
reader will find
more details about our simulation.
The parameters of the lattices studied in this work are listed
in
table~\ref{table_parameter}.

The pseudoscalar decay constant $f_P$ is extracted from the
lattice matrix
element through the relation
\be
   <0|{\cal M}_{\gamma_4 \gamma_5}^{loc}|P>
        =     Z_A^{-1}\sqrt{M_P/2} \> f_P \> a^{3/2}\;\;.
\label{deffp}
\ee
$Z_A$ is the axial current renormalisation constant,
${\cal M}_{\gamma_4 \gamma_5}^{loc}$ is the fourth component
of the local axial vector current,
and $|P>$ denotes the pseudoscalar ground state. The precise
normalization is further discussed in section 3.
We define a generalized lattice current  by
\be
   {\cal M}^{J}_{\Gamma}(\vx,t) = \bar{h}(\vx,t) \> \Gamma \>
l^J(\vx,t)
\ee
where
\be
   l^J(\vx,t)=\sum_{\vy} \Phi^J(\vx,\vy,{\cal U}(t)) \> l(\vy,t)
\ee
is a smeared light ($l$) quark field obtained by applying the
trial
wave  function $\Phi^J$ and $h(\vx,t)$ the local heavy ($h$)
quark field.\footnote{Since only smeared light quark
fields are needed for the present investigation we suppress the
source index
for the heavy quark field which was used in ref.~\cite{fb3}.}

The aim is to extract the local matrix element by using wave
functions
optimized to yield early ground state dominance in the
meson-meson
correlator
\be
   C_{\Gamma}^{I,J}(t)= \sum_\vx < {\cal M}_\Gamma^I(\vx,t)
                 \> [{\cal M}_\Gamma^J(\v0,0)]^\dagger >  \;\;.
\ee
We consider  four  methods  to extract the {\it local} matrix
element
$<0|{\cal M}_{\gamma_4 \gamma_5}^{loc}|P>$ from the local-smeared
and
smeared-smeared correlators:
\begin{enumerate}
\item[(a)]
A three-parameter simultaneous  fit to
$C^{loc,J}_{\gamma_4\gamma_5}(t)$ and
$C^{J,J}_{\gamma_4\gamma_5}(t)$
with equal  mass in both correlators.
\item[(b)]
Same as (a), but with correlators $C^{J,J}_{\gamma_5}(t)$ and
$\sum_{\vx}<{\cal M}_{\gamma_4\gamma_5}^{loc}(\vx,t) \>
   [{\cal M}_{\gamma_5}^J(\v0,0)]^\dagger>$.
\item[(c)]
A constrained fit to the local-smeared correlator
$C_{\gamma_4\gamma_5}^{loc,J}(t)$, using the mass extracted from
the
smeared-smeared correlator, $C_{\gamma_4\gamma_5}^{J,J}(t)$.
\item [(d)]
A fit to the ratio
\be
   R(t) =
      \frac{C_{\gamma_4\gamma_5}^{loc,J}(t)}
           {\sqrt{C_{\gamma_4\gamma_5}^{J,J}(t)}}
      \;\; \stackrel {t \>\mbox{\scriptsize large}} {\ra} \;\;
      <0|{\cal M}^{loc}_{\gamma_4\gamma_5}|P>  \; e^{-M_Pt/2}
\;\;.
\ee
\end{enumerate}
All fits (unless stated otherwise) have been carried out with the
inverse covariance matrix but we have checked that the mean values
are not significantly different from the ones resulting from a diagonal
$\chi^2$ fit.
Ground state dominance has been monitored both by $\chi^2$ and
by
the plateau in the local  mass
\be
\mu_{\Gamma}^{I,J}=\ln\frac{C_{\Gamma}^{I,J}(t)}{C_{\Gamma}^{I
,J}(t-a)}   \quad \mbox{in cases (a) -- (c)}
\ee
or
\be
   \mu_{R}(t) = \ln\frac{R(t)}{R(t-a)} \quad \mbox{in case (d)}.
\label{mloc}
\ee

In table~\ref{table_diff_methods}, we show the results of the
different methods
for various heavy-light quark combinations. Methods (a) -- (d)
are  seen to yield compatible results once ground state dominance
is established by a low $\chi^2$ value.

For the subsequent analysis we chose to apply method (d). The
quality of the
ground state dominance achieved is demonstrated in
figure~\ref{figure_local_mass} which shows
the local masses  for $\beta=$5.74, 6.00 and 6.26 at a light
quark mass of about $2m_s$. We list our results for
$f_P$ and $M_P$ at several combinations of
quark masses in table~\ref{table_fit_results}.


\subsection{How good are local operators?}

Having established the success of our smearing techniques we are
in
the position to examine the onset of ground state dominance in
the
case of purely local correlators.
The ratio of smeared and local correlators\cite{Factor}
\be
   r_{\Gamma}(t)=\frac{C_{\Gamma}^{J,J}(t) \;\;
      C_{\Gamma}^{loc,loc}(t)}{(C_{\Gamma}^{loc,J}(t))^2}
\label{ratio}
\ee
can be expressed in terms of completely local quantities for
sufficiently large time separations~$t$,
\be
   r_{\Gamma}^{as}(t)=
      \left( 1+\alpha_{loc}^2 \;
              \frac {e^{-(m+\Delta)t} + e^{-(m+\Delta)(T-t)}}
                    {e^{-mt} + e^{-m(T-t)}}  \right),
\label{asratio}
\ee
if the correlators are dominated by two states with energies $m$
and $m+\Delta$:
\bes
   C^{I,J}_{\Gamma}&=&A^0_I \; A^0_J (e^{-mt} + e^{-m(T-t)})
      + A^1_I \; A^1_J (e^{-(m+\Delta)t} +
e^{-(m+\Delta)(T-t)})\\ \nonumber
A_I^n &= &<0|{\cal M}^I_{\gamma_4\gamma_5}|P,n>
\ees
and as long as smearing has been operative in the sense that it
has suppressed the excited state significantly:
\be
\alpha_J \equiv A_J^1/A_J^0 \ll \alpha_{loc} \equiv A_{loc}^1/A_{loc}^0~.
\label{alpha_ineq}
\ee
The merits of this factorization test are the following:
\begin{itemize}
\item up to terms of order $\alpha_J^2$ the ratio is purely dependent
on the
      local parameters,
\item the realm of ground state dominance can be clearly
identified
      since the height of the plateau is known to be one,
\item the ratio is dimensionless and has a continuum limit in the
range where
eq. \ref{asratio} holds.
\end{itemize}

In order to determine the onset of ground state dominance in the
purely local case  we use the combined set of the present data
and
our previous local results~\cite{fb1,fb2} as an input to
eq.~\ref{ratio}. We show the ratio $r_{\gamma_4\gamma_5}$ in
figure~\ref{figure_ratio_sl} at
$\beta=6.26$ for fixed light quark mass ($m_l\simeq 2m_s$) and
different heavy quark masses as a function of the time
separation in units of GeV$^{-1}$.
It appears that the plateau regime is reached only at
about 5 GeV$^{-1}$.
The local data alone seem to allow for an earlier plateau in the
local
mass.
According to the present test, however, exploiting
the local data under the assumption of ground state dominance at
smaller time separations  leads to an
overestimation of $f_P$~\cite{fb2}.
The pre-asymptotic behaviour according to eq.~\ref{asratio} is
reflected by the data with a mass gap in the region of 600 MeV
as
illustrated by the dashed lines in figure~\ref{figure_ratio_sl}.

At all values of the heavy quark mass that we investigated, the
gap is consistent with $\sim 600$~MeV. The amplitude  $\alpha_{loc}$
does, however, increase significantly when the mass of the heavy quark
is increased. This entails that the {\it mass--dependence} of the decay
constant is affected by fitting too early. On the basis of the present
data,
we estimate that this effect is of the order of the statistical
errors for the $\beta=6.4$ data of ref.~\cite{Abada} as well
as for the $\beta=6.26$ results of ref.~\cite{fb2}.

In the above interpretation we rely on eq.~\ref{alpha_ineq}.
To some extent this equation is
checked by the lattice spacing
independence of $r_{\Gamma}$ which results from eq.~\ref{alpha_ineq}
(of course, as for any observable that has a continuum limit,
the normal kind of $a$--effects should be there).
In
figure~\ref{figure_ratio_with_beta} we
display $r_{\Gamma}$ for different values of $\beta$. In
agreement with eq.~\ref{alpha_ineq}, we find that
$r_{\Gamma}$ is independent of $\beta$
within the statistical errors.


\section{Renormalization}

Our quark fields are normalized according to the relativistic
normalization (see ref.~\cite{fb3} for the form of the action).
Due to the broken chiral symmetry in the Wilson formulation,
the lattice axial vector current needs a finite renormalization~\cite{KaSm}
\be
A_\mu=Z_A {\cal M}_{\gamma_{\mu} \gamma_5}^{loc}~.
\ee

As an intrinsic short distance
quantity, $Z_A$  can be calculated in perturbation theory,
which  at one-loop order
gives~\cite{ZA}
\be
   Z_A=1-0.1333 \:g^2 \;\;.
\label{ZAsta}
\ee
Since the next order term is so far unknown, it is not {\it a priori}
clear which coupling constant $g$ should be used to evaluate $Z_A$
numerically. It is known that the bare coupling $g_0$ is not a good
expansion parameter~\cite{LM}.
A recent analysis of a number of short
distance
dominated quantities by Lepage and Mackenzie~\cite{LM} revealed
that a mean
field approach~\cite{MFT}, which absorbs lattice tadpole
contributions
into effective quantities in the Wilson action, leads to a
substantially better
agreement between lattice Monte Carlo results and their
perturbative expansions. We therefore use in the following the
mean field improved coupling
\be
  \tilde{g}^2=g_0^2/P \;\; {\rm \ with \ } P=<\frac{1}{3}\tr P_{\mu\nu}> \;,
\ee
as determined from the Monte Carlo simulation.
In detail, also other forms of mean field improvements are being
used~\cite{LM,BLSnew},
but we note that they do not differ appreciably in numerical values and
{\it  we have to remember that the errors remain of
order $O(\tilde{g}^4)$} in any case.

Kronfeld and
Mackenzie~\cite{KM} have argued that for quark masses that are
comparable to the inverse lattice spacing, an -- at order
$O(a m_q)$ -- different normalization of the axial vector current
should have matrix elements with smaller lattice artifacts.
For comparison, we use this nonrelativistic normalization
\be
\tilde{A}_\mu=\tilde{Z}_A \sqrt{\frac{1}{2\kappa_l} - \frac{3}{8\kappa_c}}
                  \sqrt{\frac{1}{2\kappa_h} - \frac{3}{8\kappa_c}}
{\cal M}_{\gamma_{\mu} \gamma_5}^{loc}~, \label{kronfeld}
\ee
with $l$ and $h$ labelling the light and the heavy quark field.
Here,
$\kappa_c$ denotes the critical value of the hopping parameter as determined
from the vanishing of the pion mass.
The renormalization constant $\tilde{Z}_A$ is given by~\cite{LM}
\be
   \tilde{Z}_A=1-0.0248\: \tilde{g}^2\;\;.
\ee
In the following section we will use both $A_{\mu}$ and
$\tilde{A}_\mu$
for analyzing the data. The latter is denoted by `{\KroMac}
norm'.


\section{Results}

\subsection{Finite $a$ effects}

In order to determine physical quantities from lattice
calculations, it is crucial to perform the
extrapolation $a \ra 0$, within a chosen scale.
In principle one would like to use $f_\pi$ for this purpose, since
then the $O(\tilde{g}^4)$ uncertainty in $Z_A$ cancels.
At present, the statistical errors of $f_\pi$ in our simulation are
however too large to pursue this approach.
Therefore we will use the string tension, which has small statistical
errors,  during the extrapolation
and convert the result  finally into the most `natural' scale,
$f_\pi$.
As stated in the introduction, the supposed improvement of the
normalization ( eq.~\ref{kronfeld} )
should manifest itself in  a smaller slope of $f_P(a)$
as $a$ goes to zero~\cite{KM}.
 The question is whether this is actually
the
case
in the available range of $\beta$. In order to check this we have
computed $f_P$ both with the naive (standard relativistic) and the
{\KroMac} norms at several  pseudoscalar masses. The data from
table~\ref{table_fit_results} are
extrapolated in the standard manner
to the chiral limit (see ref.~\cite{fb3}) and are compiled in
table~\ref{table_extra_results}.
We have determined the
$a$-dependence at fixed values of $M_P$ within the range 1.1 GeV
$\le
M_P \le$ 2.3 GeV. In figure~\ref{figure_fp_extra_a_sigma} we show
the
situation with relativistic norm and with the {\KroMac} norm.
Note, that we had to interpolate between the  computed $M_P$
values  in order to
tune for fixed physical $M_P$ as $\beta$ is varied.
The interpolation introduces only a minor uncertainty since the
dependence of $f_P$ on $M_P$ is very weak. It was
estimated by comparing two--point and three--point interpolations
and added to the statistical errors.

It can be
seen that the {\KroMac} norm has a sizeable
impact
on $f_P$ at present values of the lattice spacing.
Contrary to the expectations,  however, $f_P(a)$ is
turned from a
very weakly decreasing into a strongly decreasing function.
As we increase $M_P$ the effect becomes more pronounced.
Nevertheless the extrapolated values at $a=0$ coincide.
 We have used a linear extrapolation and omitted those
points
where the effect of the {\KroMac} factor amounts to more than a 60
\% change. The linear extrapolation functions have been
followed back to the
omitted points as
dashed lines. These demonstrate that the extrapolation is stable,
namely, within the statistical
errors one would obtain the same result at
$a=0$ if all points were included in the extrapolation.
Note that according to perturbative arguments, the leading lattice
spacing dependence should be linear in $a$, {\it if} the
$O(\tilde{g}^4)$ terms are numerically small. We come back to that
point in section~4.3.

In figure~\ref{figure_fp_v_mp_inv} we display the effects of the
{\KroMac}
factor onto the quantity
\be
   \fhat_P = f_P \sqrt{M_P}
       \left (\frac{\alpha_s(M_P)}{\alpha_s(M_B)}
\right)^{6/33}\; .
\label{fhat}
\ee
The first impression is that
at {\em fixed} and {\em finite} values of $a$ the
data in {\KroMac} norm  interpolate smoothly
between the
$D$~region and the static point and this has been interpreted as
improvement~\cite{BLSnew,UKQCD}.
 Given the conclusion from
figure~\ref{figure_fp_extra_a_sigma} the improvement does not
survive
the limit $a \ra 0$, however. This is visualized in
figure~\ref{figure_fp_v_mp_inv} by the error band which refers
to the
results of the above continuum extrapolation.
Note also, that the poor  scaling behavior of $f_P$ in the {\KroMac} norm
is already visible in figure~5.18 of ref.~\cite{BLSnew}.

\subsection{Finite size effects}

The data we discussed above was obtained on lattices with periodicity in
space $L$ of magnitude $L\sqrt{\sigma}\sim 3$.
Here  we intend to consolidate our results with respect to
possible  finite size effects. For that purpose we have plotted
in figure~\ref{figure_fp_volume} $f_P$ as a function of $L$
for three pseudoscalars with fixed light quark mass of about
2$m_s$ at
$\beta=5.74$.
 As expected finite size
effects decrease with increasing  heavy quark mass.

In the range $L\sqrt{\sigma}\geq 3$ we find no significant finite
volume effects with a precision of the order of $\sim 2 \%$.
Decreasing the light quark mass down to $m_s$ we compare
the $8^3 \times 24$ and $12^3
\times
24$ lattices at $\beta=5.74$, c.f. table~\ref{table_fit_results}.
There appears to be a small
 volume effect of maximally $3\%\pm 2\%$.
This is, however, statistically not significant.
So we have kept  the
volume
fixed at $L\sqrt{\sigma}\sim 3$ throughout our final analysis,
but estimate from the above analysis, that there may be a $\sim 5$\%
error involved with this.

\subsection{Conversion to scale $f_{\pi}$ }

The most 'natural' scale for $f_P$ is
$f_{\pi}$,
since the uncertainty due to  the renormalization
constant $Z_A$ cancels out in this
case and one may also hope, that the ratio $f_P / f_{\pi}$
is not affected strongly by the quenched approximation.
Lattice measurements of $f_{\pi}$ are generally accompanied by
large statistical errors and therefore
we have decided  to convert our results to this scale  only after
having performed the
$a \rightarrow 0$ extrapolation of $f_P$.  To achieve this we
have decoupled the extrapolations
according to
\be
   \frac{f_P}{f_{\pi}}(a \rightarrow 0) =
      \frac{{f_P}/{\sqrt{\sigma}}(a \rightarrow 0)}
           {{f_{\pi}}/{\sqrt{\sigma}}(a \rightarrow 0)}\quad .\label{conrat}
\label{convert}
\ee
To obtain the  denominator of eq.~\ref{convert} we used
both  our own data and  the results of other
lattice groups~\cite{BLSnew,Ape,UKQCD_light,GF11} (all data
has been changed to the
relativistic normalization).
The compilation  of the data is shown in
figure~\ref{figure_fpi_sigma}.
Since the $a$ dependence of ${f_{\pi}}/{\sqrt{\sigma}}$ is
obviously
weak, a linear extrapolation to $a=0$ is well justified and leads
to :
\be
   \frac{f_{\pi}}{\sqrt{\sigma}}(a=0) = 0.269(12) \quad .
\ee
In addition to the quoted statistical error, this quantity has an
uncertainty due to the missing $O(\tilde{g}^4)$ terms in $Z_A$.
This should, however, largely cancel out when we take the ratio
eq.~\ref{conrat}.

\subsection{Heavy mass extrapolation}

In figure~\ref{figure_fp_extra_mh_fpi} we display our final
results\footnote{To be specific the extrapolation has been
performed on the
data using the relativistic norm since it involves smaller
statistical
errors than
using the {\KroMac} norm.}  at $a = 0$ in the
form $\hat{f}_P(1/M_P)$, together with our static value from
ref.~\cite{fb3}.
As described in the previous section the scale has finally been
converted to
$f_{\pi}=132$MeV.
The new data appears to depend only weakly on $M_P$.
Note however, that the data points carry
error bars of order $25 \%$ and
therefore do not exclude a stronger variation in $M_P$.
This is due to the
  extrapolations to the  chiral and continuum limits.

Given this  situation  we draw an error band that links the
conventional
results with the static point. The $M_P$ dependence of the error
band was
chosen according to the ansatz
\be
   \hat{f}_P = c_0 +\frac{c_1}{M_P} + \frac{c_2}{M_P^2}\quad .
\ee
At the location of the B  meson the error band corresponds
to
\be
   f_B = 0.18(5) \mbox{GeV}  \quad \quad.
\label{result1}
\ee
It is evident from figure~\ref{figure_fp_extra_mh_fpi} that this
value
is strongly affected by the size and uncertainty of
$f_{\mbox{stat}}$.
{}From the direct continuum extrapolation at the mass of the $D$, we quote
\be
   f_D = 0.17(3) \mbox{GeV}  \quad \quad.
\label{result2}
\ee

At a given $M_P$ one can extrapolate to $a \ra 0$ the ratio
$f_{P_s}/f_{P}$. In the ratio scale ambiguities drop
out and additional systematic errors e.g. due to quenching may
be reduced. We find
that the heavy mass dependence of the ratio is very weak and can
be taken as
constant in the mass range between the charm and the B mesons.
In the continuum
limit we obtain
$
   f_{P_s}/f_{P}=1.09(2)
$ at $L\sqrt{\sigma} \sim 3$, where the statistical error is quite small.
We therefore have to remember the finite size uncertainty and finally
quote
$$
   f_{P_s}/f_{P}=1.09(2)(5).
$$


\section{Summary and conclusions}
We have carried out a high statistics analysis of the
pseudoscalar decay
constant. In the case of purely local operators, the ground state
signal is
attained only at time separations larger than 5 GeV$^{-1}$.
For this reason smearing techniques should also be applied  in
the nonstatic
situation.

We observe little volume dependence of $f_P$ as long as
$L\sqrt{\sigma}
\geq 3$.
 For  $5.74 \le \beta \le 6.26$, we discover
a marked variation with $a$,  if the {\KroMac}
normalization is used, while the relativistic
normalization results in
$f_P$-values that are rather insensitive to our variation of $a$.
This implies that {\KroMac}  does not yet lead to improvement in the
sense that
finite $a$ results are shifted  towards their continuum limit.

In this context, we emphasize that it  is  misleading
to look only at the $M_P$ dependence of $f_P$ at one value of
$\beta$.
One rather has to work at different $\beta$ values and to
extrapolate
to the continuum. Proceeding in this manner we find that the
difference originating from
the two normalizations vanishes, as it should.

We consider the present investigation as exploratory
in spite of the statistics of about 100
independent configurations on the largest lattice.
The estimates of  $f_B$ and $f_D$
(see eqs.~\ref{result1},\ref{result2}) still
carry  rather large errors originating from the
unavoidable extrapolation in $a$. Study of the $a$
dependence is also necessary when an improved action is used.
The extrapolation errors can be greatly reduced
  with the power of available parallel
supercomputers,
by increasing  the number of points in $a$  with
  very high statistics.

As we have mentioned, it would also be desirable to perform the
whole continuum extrapolation on the ratio $f_P/f_\pi$ in order
to be completely free of the uncertainties in $Z_A$.
It appears, however, that better  ways of computing $f_\pi$ need
to be found before this is feasible.

Along the same lines, a study of the finite $a$ effects of
$B_B$ is necessary, since it has
been clearly demonstrated for the case of $B_K$ that $a$ effects
do not necessarily cancel out for these observables~\cite{Sharp}.

{\bf Acknowledgements.} The calculations were carried out on the
NEC-SX3 at the Supercomputing
Center in Manno, Switzerland and on the CRAY-Y-MP of the HLRZ in
J\"ulich, Germany.
We thank both institutions and their staff for their generous
support.

\clearpage
\pagestyle{empty}
\begin{table}[p]
{\Large \bf Tables }\\
\vspace{1ex}
\caption{\protect{\hspace*{14cm}} \label{table_parameter}}
\vspace{1ex}

Pseudoscalar decay constant and mass in lattice units as a
function of $\tmin$, the smallest time
separation used in the fit. Smearing is used
for the light quark propagator. The results correspond to a light
quark
mass of about twice the strange quark mass (or $M(l,l)^2/\sigma
\sim
4$). $\chi^2$ denotes $\chidof$.

\vspace{1ex}
\vspace{1ex}

{\small
  \begin{center}
    \fbox{
      \begin{minipage}{14cm}
        \center{$\beta=5.74 \quad \kappa_1=0.1560 \quad \kappa_2=0.1250$}\\
        \begin{tabular}{|c|c|c|c|}
          \multicolumn{4}{c}{$4^3 \times 24 \quad$ 404 conf.}\\
          \hline
          $\tmin$ & $af/Z_A$ & $aM$ & $\chi^2$ \\
          \hline
           3 & 0.1954(39) & 1.334(13) & uncorr. \\
          \hline
          \multicolumn{4}{c}{} \\
          \multicolumn{4}{c}{$6^3 \times 24 \quad$ 131 conf.} \\
          \hline
          $\tmin$ & $af/Z_A$ & $aM$ & $\chi^2$ \\
          \hline
           3 & 0.2159(18) & 1.346(5)  & uncorr. \\
          \hline
          \multicolumn{4}{c}{} \\
          \multicolumn{4}{c}{$8^3 \times 24 \quad$ 175 conf.}\\
          \hline
          $\tmin$ & $af/Z_A$ & $aM$ & $\chi^2$ \\
          \hline
           2 & 0.2270(19) & 1.360(7) & 3.1 \\
           3 & 0.2250(19) & 1.359(6) & 3.1 \\
           4 & 0.2222(22) & 1.356(7) & 2.5 \\
           5 & 0.2207(28) & 1.355(5) & 2.7 \\
           6 & 0.2169(35) & 1.350(4) & 2.6 \\
           7 & 0.2157(38) & 1.350(6) & 2.5 \\
           8 & 0.2209(43) & 1.353(8) & 2.3 \\
          \hline
        \end{tabular}
        \hfill
        \begin{tabular}{|c|c|c|c|}
          \multicolumn{4}{c}{$10^3 \times 24 \quad$ 213 conf.}\\
          \hline
          $\tmin$ & $af/Z_A$ & $aM$ & $\chi^2$\\
          \hline
           2 & 0.2257(9)  & 1.353(4) & 2.9 \\
           3 & 0.2235(4)  & 1.353(5) & 2.3 \\
           4 & 0.2206(16) & 1.351(4) & 1.5 \\
           5 & 0.2193(18) & 1.350(3) & 1.4 \\
           6 & 0.2166(18) & 1.347(4) & 0.4 \\
           7 & 0.2161(22) & 1.347(5) & 0.4 \\
           8 & 0.2148(32) & 1.346(3) & 0.3 \\
          \hline
          \multicolumn{4}{c}{} \\
          \multicolumn{4}{c}{$12^3 \times 24 \quad$ 113 conf.}\\
          \hline
          $\tmin$ & $af/Z_A$ & $aM$ & $\chi^2$\\
          \hline
           2 & 0.2212(4)  & 1.356(12)& 1.0 \\
           3 & 0.2208(23) & 1.355(3) & 1.3 \\
           4 & 0.2181(20) & 1.353(9) & 0.6 \\
           5 & 0.2175(30) & 1.353(3) & 0.6 \\
           6 & 0.2172(28) & 1.353(6) & 0.8 \\
           7 & 0.2157(30) & 1.351(7) & 0.7 \\
           8 & 0.2153(36) & 1.351(4) & 1.0 \\
          \hline
        \end{tabular}
      \end{minipage}
   } 
 \end{center}
\fbox{
  \begin{tabular}{|c|c|c|c|}
    \multicolumn{4}{c}{$\beta=6.00\quad\kappa_1=0.1525\quad\kappa_2=0.1250$}\\
    \multicolumn{4}{c}{$12^3 \times 36 \quad$ 204 conf.} \\
    \hline
    $\tmin$ & $af/Z_A$ & $aM$ & $\chi^2$ \\
    \hline
     3 & 0.1280(14) & 1.080(2) & 2.0 \\
     4 & 0.1277(14) & 1.079(2) & 2.0 \\
     5 & 0.1260(15) & 1.078(3) & 1.0 \\
     6 & 0.1266(15) & 1.078(3) & 1.1 \\
     7 & 0.1273(19) & 1.079(3) & 1.1 \\
     9 & 0.1270(19) & 1.079(4) & 1.4 \\
    \hline
    \multicolumn{4}{c}{} \\
    \multicolumn{4}{c}{$18^3 \times 36 \quad$ 9 conf.} \\
    \hline
    $\tmin$ & $af/Z_A$ & $aM$ & $\chi^2$ \\
    \hline
     6 & 0.1281(54) &1.084(7) & uncorr. \\
     7 & 0.1264(57) &1.082(7) & uncorr. \\
     9 & 0.1208(76) &1.075(11)& uncorr. \\
    \hline
  \end{tabular}
}
\hfill
\fbox{
  \begin{tabular}{|c|c|c|c|}
    \multicolumn{4}{c}{$\beta=6.26\quad
\kappa_1=0.1492\quad\kappa_2=0.1200$}\\
    \multicolumn{4}{c}{$12^3 \times 48 \quad$ 103 conf.} \\
    \hline
    $\tmin$ & $af/Z_A$ & $aM$ & $\chi^2$ \\
    \hline
     4 & 0.0859(23) & 1.040(5) & 1.8 \\
     6 & 0.0845(34) & 1.039(8) & 2.0 \\
     8 & 0.0834(30) & 1.039(6) & 1.6 \\
     10& 0.0853(46) & 1.039(7) & 1.2 \\
     12& 0.0831(47) & 1.035(7) & 1.2 \\
     14& 0.0808(48) & 1.031(9) & 1.2 \\
     16& 0.0823(55) & 1.033(8) & 1.5 \\
    \hline
    \multicolumn{4}{c}{} \\
    \multicolumn{4}{c}{$18^3 \times 48 \quad$ 76 conf.} \\
    \hline
    $\tmin$ & $af/Z_A$ & $aM$ & $\chi^2$ \\
    \hline
     4  & 0.0833(21) & 1.039(7)& 2.9 \\
     6  & 0.0813(21) & 1.034(6)& 2.0 \\
     8  & 0.0811(27) & 1.034(5)& 2.3 \\
     10 & 0.0784(29) & 1.030(5)& 1.3 \\
     12 & 0.0777(29) & 1.029(5)& 1.5 \\
     14 & 0.0805(39) & 1.033(6)& 1.5 \\
     16 & 0.0783(44) & 1.030(7)& 1.7 \\
    \hline
  \end{tabular}
} 
} 
\end{table}

\clearpage
\begin{table}[p]
\caption{\protect{\hspace*{14cm}} \label{table_diff_methods}}
\vspace{1ex}

$a f/Z_A$ calculated with different methods as described
in the text: \\
(a) simultaneous fit to $C_{\gamma_4\gamma_5}^{J,J}$
and $C_{\gamma_4\gamma_5}^{loc,J}$, \\
(b) simultaneous fit to $C_{\gamma_5}^{J,J}$ and
$\sum_{\vx} <M_{\gamma_4\gamma_5}^{loc}({\vx},t)
[M_{\gamma_5}^{J}(\vec{0},0)]^\dagger>$, \\
(c) determine $a\:m_P$ by fit to
$C_{\gamma_4\gamma_5}^{J,J}$
and use this as constraint in the fit to
$C_{\gamma_4\gamma_5}^{loc,J}$, \\
(d) ratio method.

We show the results for the lattices used to obtain our final
results. $\chi^2$ actually denotes $\chidof$.

\begin{center}
\begin{tabular}{|c|cc|cc|cc|cc|}
\multicolumn{9}{c}{$\beta=5.74\quad N_S^3 \times N_T = 8^3 \times
24
   \quad \kappa_1 = 0.1560 \quad \kappa_2 = 0.1250 $ }\\
\multicolumn{9}{c}{} \\
\hline
$\tmin$ & (a) & $\chi^2$ & (b) & $\chi^2$ & (c) & $\chi^2$ & (d)
& $\chi^2$ \\
\hline
2 & 0.2274(14) & 2.8 & 0.2372(26) & 8.8 & 0.2314(17) & 6.8 &
0.2270(19) & 3.1\\
3 & 0.2256(20) & 2.6 & 0.2313(28) & 6.9 & 0.2280(23) & 3.0 &
0.2250(19) & 3.1\\
4 & 0.2224(24) & 2.1 & 0.2250(28) & 4.1 & 0.2238(28) & 2.2 &
0.2222(22) & 2.5\\
5 & 0.2203(27) & 1.9 & 0.2216(29) & 3.4 & 0.2196(30) & 1.6 &
0.2207(28) & 2.7\\
6 & 0.2159(35) & 1.7 & 0.2158(31) & 1.4 & 0.2169(33) & 2.0 &
0.2169(35) & 2.6\\
7 & 0.2145(41) & 2.0 & 0.2142(35) & 1.4 & 0.2154(36) & 2.6 &
0.2157(38) & 2.5\\
8 & 0.2181(44) & 1.7 & 0.2130(38) & 1.8 & 0.2144(38) & 1.3 &
0.2209(43) & 2.3\\
\hline
\multicolumn{9}{c}{} \\
\multicolumn{9}{c}{$\beta=6.00\quad N_S^3 \times N_T = 12^3
\times 36 \quad
\kappa_1 = 0.1525 \quad \kappa_2 = 0.1250$} \\
\multicolumn{9}{c}{} \\
\hline
$\tmin$ & (a) & $\chi^2$ & (b) & $\chi^2$ & (c) & $\chi^2$ & (d)
& $\chi^2$ \\
\hline
3 & 0.1267(14) & 1.8 & 0.1333(14) & 4.9 & 0.1309(16) & 1.8 &
0.1280(17) & 2.0\\
4 & 0.1266(10) & 1.9 & 0.1318(15) & 3.5 & 0.1303(16) & 1.8 &
0.1277(14) & 2.0\\
5 & 0.1251(14) & 1.2 & 0.1307(15) & 1.7 & 0.1268(20) & 1.4 &
0.1260(15) & 1.0\\
6 & 0.1257(17) & 1.1 & 0.1296(18) & 1.7 & 0.1281(16) & 1.3 &
0.1266(15) & 1.1\\
7 & 0.1262(22) & 1.2 & 0.1282(20) & 1.5 & 0.1286(20) & 1.5 &
0.1273(19) & 1.1\\
9 & 0.1248(20) & 1.3 & 0.1257(23) & 1.1 & 0.1257(26) & 1.8 &
0.1270(19) & 1.4\\
11& 0.1236(26) & 1.4 & 0.1243(23) & 1.1 & 0.1231(30) & 2.0 &
0.1258(24) & 1.7\\
\hline
\multicolumn{9}{c}{} \\
\multicolumn{9}{c}{$\beta=6.26\quad N_S^3 \times N_T = 18^3
\times 48 \quad
\kappa_1 = 0.1492 \quad \kappa_2 = 0.1200$} \\
\multicolumn{9}{c}{} \\
\hline
$\tmin$ & (a) & $\chi^2$ & (b) & $\chi^2$ & (c) & $\chi^2$ & (d)
& $\chi^2$ \\
\hline
4 & 0.0819(30) & 2.5 & 0.0842(42) & 3.3 & 0.0821(36) & 2.2 &
0.0833(21) & 2.9\\
6 & 0.0809(33) & 2.6 & 0.0825(39) & 2.7 & 0.0810(36) & 2.3 &
0.0813(21) & 2.0\\
8 & 0.0799(30) & 2.6 & 0.0808(32) & 2.8 & 0.0811(37) & 2.4 &
0.0811(27) & 2.3\\
10& 0.0770(25) & 1.6 & 0.0782(22) & 1.3 & 0.0777(33) & 1.4 &
0.0784(29) & 1.3\\
12& 0.0775(30) & 1.6 & 0.0783(24) & 1.2 & 0.0782(38) & 1.7 &
0.0777(29) & 1.5\\
14& 0.0800(36) & 1.5 & 0.0795(29) & 1.1 & 0.0800(34) & 1.8 &
0.0805(39) & 1.5\\
16& 0.0789(35) & 1.5 & 0.0793(29) & 1.2 & 0.0782(41) & 1.7 &
0.0783(44) & 1.7\\
\hline
\end{tabular}
\end{center}
\end{table}

\clearpage
\begin{table}[p]
\caption{\protect{\hspace*{14cm}} \label{table_fit_results}}
\vspace{1ex}

Pseudoscalar decay constant and mass in lattice units. The data has
been used in the detailed analysis to extract $f_B$ and
$f_D$, with the exception of the results at $\beta=5.74$, $12^3 \times 24$.
 Smearing has been applied to the light quark propagator.
$\kappa_1$
corresponds to the light quark. $\chi^2$ actually denotes
$\chidof$.

\begin{center}
\begin{tabular}{|cc|c|c|c|c|}
\multicolumn{6}{c}{$\beta=5.74\quad N_S^3 \times N_T = 8^3 \times
24 $}\\
\hline
$\kappa_1$ & $\kappa_2$ & fit range & $a\>M$ & $a\>f/Z_{A}$ &
$\chi^2$ \\
\hline
0.156 & 0.09  & 5--12 & 1.999(4) & 0.1905(33) & 1.0 \\
      & 0.125 & 4--12 & 1.344(4) & 0.2206(24) & 2.8 \\
      & 0.140 & 4--12 & 1.065(4) & 0.2231(31) & 3.0 \\
      & 0.150 & 4--12 & 0.870(4) & 0.2164(41) & 2.4 \\
      & 0.156 & 4--12 & 0.748(5) & 0.2008(57) & 2.4 \\
\hline
0.162 & 0.09  & 5--12 & 1.924(8) & 0.1750(40) & 1.0 \\
      & 0.125 & 4--12 & 1.261(2) & 0.2022(23) & 1.8 \\
      & 0.140 & 4--12 & 0.969(5) & 0.2037(33) & 2.1 \\
      & 0.150 & 4--12 & 0.761(4) & 0.1972(48) & 2.0 \\
      & 0.162 & 5--12 & 0.474(8) & 0.1608(58) & 1.1 \\
\hline
0.1635 & 0.09  & 5--12 & 1.904(10)& 0.1697(43) & 0.8 \\
       & 0.125 & 4--12 & 1.242(7) & 0.1973(24) & 2.0 \\
       & 0.140 & 4--12 & 0.948(5) & 0.1997(30) & 1.1 \\
       & 0.150 & 4--12 & 0.734(4) & 0.1917(44) & 1.3 \\
       & 0.1635& 5--12 & 0.392(12)& 0.1479(64) & 1.1 \\
\hline
\multicolumn{6}{c}{} \\
\multicolumn{6}{c}{$\beta=5.74\quad N_S^3 \times N_T = 12^3 \times 24 $}\\
\hline
$\kappa_1$ & $\kappa_2$ & fit range & $a\>M$ & $a\>f/Z_{A}$ &
$\chi^2$ \\
\hline
0.156 & 0.09  & 5--12 & 2.001(6) & 0.1872(24) & 0.6 \\
      & 0.125 & 4--12 & 1.354(5) & 0.2179(24) & 0.8 \\
      & 0.140 & 4--12 & 1.073(7) & 0.2207(12) & 1.1 \\
      & 0.150 & 4--12 & 0.878(5) & 0.2133(23) & 1.1 \\
      & 0.156 & 4--12 & 0.755(5) & 0.1959(24) & 0.8 \\
\hline
0.162 & 0.09  & 5--12 & 1.927(17)  & 0.1689(40) & 0.8 \\
      & 0.125 & 4--12 & 1.267(7) & 0.1974(29) & 0.6 \\
      & 0.140 & 4--12 & 0.976(9) & 0.2001(24) & 1.0 \\
      & 0.150 & 4--12 & 0.768(4) & 0.1929(22) & 1.1 \\
      & 0.162 & 5--12 & 0.481(2) & 0.1628(30) & 0.8 \\
\hline
0.1635 & 0.09  & 5--12 & 1.911(15)& 0.1631(47) & 0.7 \\
       & 0.125 & 4--12 & 1.246(5) & 0.1914(27) & 0.5 \\
       & 0.140 & 4--12 & 0.953(8) & 0.1935(23) & 0.9 \\
       & 0.150 & 4--12 & 0.740(2) & 0.1864(30) & 1.1 \\
       & 0.1635& 5--12 & 0.395(4) & 0.1523(35) & 0.9 \\
\hline
\end{tabular}
\end{center}
\end{table}
\clearpage
\begin{center}
\begin{tabular}{|cc|c|c|c|c|}
\multicolumn{6}{c}{$\beta=6.00\quad N_S^3 \times N_T = 12^3
\times 36$} \\
\hline
$\kappa_1$ & $\kappa_2$ & fit range & $a\>M$ & $a\>f/Z_{A}$ &
$\chi^2$ \\
\hline
 0.1525 & 0.10   & 5--13 & 1.574(7) & 0.1060(15) & 0.7 \\
        & 0.115  & 5--13 & 1.279(3) & 0.1186(14) & 0.6 \\
        & 0.125  & 5--18 & 1.077(2) & 0.1260(15) & 1.0 \\
        & 0.135  & 5--18 & 0.870(3) & 0.1349(13) & 0.6 \\
        & 0.145  & 7--18 & 0.649(1) & 0.1301(23) & 0.7 \\
        & 0.1525 & 4--11 & 0.452(2) & 0.1090(15) & 0.6 \\
\hline
 0.1540 & 0.10   & 5--13 & 1.546(8) & 0.1003(15) & 0.7 \\
        & 0.115  & 5--13 & 1.252(5) & 0.1123(15) & 0.6 \\
        & 0.125  & 5--18 & 1.050(2) & 0.1190(13) & 1.0 \\
        & 0.135  & 5--18 & 0.840(4) & 0.1243(14) & 0.6 \\
        & 0.145  & 5--18 & 0.617(1) & 0.1211(18) & 0.9 \\
        & 0.154  & 4--11 & 0.368(3) & 0.0980(16) & 0.7 \\
\hline
 0.1558 & 0.10   & 5--13 & 1.521(8) & 0.0925(17) & 1.1 \\
        & 0.115  & 5--13 & 1.221(5) & 0.1036(16) & 1.1 \\
        & 0.125  & 5--18 & 1.018(5) & 0.1100(17) & 1.2 \\
        & 0.135  & 5--18 & 0.806(4) & 0.1149(18) & 0.9 \\
        & 0.145  & 5--18 & 0.579(4) & 0.1111(21) & 0.9 \\
        & 0.1558 & 4--11 & 0.257(8) & 0.0821(22) &     \\
\hline
\multicolumn{6}{c}{} \\
\multicolumn{6}{c}{$\beta=6.26 \quad N_S^3 \times N_T = 18^3
\times 48$}\\
\hline
$\kappa_1$ & $\kappa_2$ & fit range & $a\>M$ & $a\>f/Z_{A}$ &
$\chi^2$ \\
\hline
 0.1492 & 0.09   & 10--18 & 1.639(8) & 0.0577(26) & 0.5 \\
        & 0.10   & 10--18 & 1.433(7) & 0.0643(25) & 0.4 \\
        & 0.120  & 10--23 & 1.033(5) & 0.0784(29) & 1.3 \\
        & 0.135  & 10--23 & 0.707(4) & 0.0880(25) & 1.4 \\
        & 0.145  & 6--23  & 0.461(3) & 0.0882(19) & 1.3 \\
        & 0.1492 & 5--18  & 0.344(4) & 0.0759(33) & 1.0 \\
\hline
 0.1506 & 0.09   & 10--18 & 1.614(11) & 0.0517(28) & 0.9 \\
        & 0.10   & 10--18 & 1.409(9)  & 0.0578(26) & 0.7 \\
        & 0.120  & 10--23 & 1.003(7)  & 0.0709(29) & 1.4 \\
        & 0.135  & 10--23 & 0.676(4)  & 0.0799(24) & 1.3 \\
        & 0.145  & 6--23  & 0.424(4)  & 0.0810(21) & 1.2 \\
        & 0.1492 & 4--18  & 0.301(4)  & 0.0672(37) & 1.1 \\
        & 0.1506 & 4--18  & 0.254(5)  & 0.0642(31) & 1.0 \\
\hline
 0.1514 & 0.09   & 10--18 & 1.596(14) & 0.0478(29) & 0.8 \\
        & 0.10   & 10--18 & 1.391(11) & 0.0532(28) & 0.7 \\
        & 0.120  & 10--23 & 0.985(9)  & 0.0662(30) & 1.3 \\
        & 0.135  & 10--23 & 0.658(4)  & 0.0745(26) & 1.3 \\
        & 0.145  & 6--18  & 0.407(4)  & 0.0761(21) & 1.1 \\
        & 0.1492 & 4--18  & 0.275(4)  & 0.0638(39) & 1.2 \\
        & 0.1506 & 4--17  & 0.224(5)  & 0.0596(41) & 0.9 \\
        & 0.1514 & 4--17  & 0.189(5)  & 0.0565(39) & 0.7 \\
\hline
\end{tabular}
\end{center}

\clearpage
\begin{table}[p]
\caption{\protect{\hspace*{14cm}} \label{table_extra_results}}
\vspace{1ex}

We give the pseudoscalar decay constant and mass in lattice units
extrapolated to $\kappa_u$ and $\kappa_s$, as well as the ratio
of the decay constant for a light quark fixed to the  strange
quark
mass to that evaluated at the chiral limit, for all the heavy
quarks
$\kappa_h$ considered. The subscript $u$ and $s$ denote
quantities evaluated
at the chiral limit and at the strange quark mass respectively.
The strange quark mass was fixed using the $\sigma$ scale.

\begin{center}
\begin{tabular}{|c|c|c|c|c|c|}
\multicolumn{6}{c}{$\beta=5.74\quad \kappa_c=0.1664(5)\quad
                                    \kappa_s=0.1589(3)$}\\
\hline
$\kappa_h$ & $a\>f_u/Z_A$ & $a\>M_u$ & $a\>f_s/Z_{A}$ &
                       $a\>M_s$ & $f_s/f_u$ \\
\hline
 0.06  & 0.1197(102)& 2.502(13)& 0.1313(74) & 2.542(9) &
1.097(33) \\
 0.09  & 0.1629(52) & 1.871(13)& 0.1726(39) & 1.916(9) &
1.059(11) \\
 0.125 & 0.1890(33) & 1.205(7) & 0.2001(32) & 1.253(2) &
1.058(10) \\
 0.140 & 0.1907(38) & 0.904(6) & 0.2020(29) & 0.960(4) &
1.059(10) \\
 0.150 & 0.1829(53) & 0.684(5) & 0.1947(42) & 0.749(4) &
1.065(10) \\
\hline
\multicolumn{6}{c}{} \\
\multicolumn{6}{c}{$\beta=6.00\quad \kappa_c=0.1572(4)\quad
                                    \kappa_s=0.1540(3)$}\\
\hline
$\kappa_h$ & $a\>f_u/Z_A$ & $a\>M_u$ & $a\>f_s/Z_{A}$ &
                       $a\>M_s$ & $f_s/f_u$ \\
\hline

 0.10  & 0.0873(17) & 1.498(10)& 0.0936(16) & 1.523(9) & 1.073(1)

\\
 0.115 & 0.0983(18) & 1.197(7) & 0.1048(16) & 1.225(4) & 1.073(5)

\\
 0.125 & 0.1038(18) & 0.995(5) & 0.1113(16) & 1.026(2) & 1.073(6)

\\
 0.135 & 0.1085(19) & 0.780(7) & 0.1163(16) & 0.810(5) & 1.072(3)

\\
 0.145 & 0.1032(31) & 0.551(2) & 0.1123(20) & 0.584(4) &
1.088(14) \\
\hline
\multicolumn{6}{c}{} \\
\multicolumn{6}{c}{$\beta=6.26\quad \kappa_c=0.1524(6)\quad
                                    \kappa_s=0.1507(4)$}\\
\hline
$\kappa_h$ & $a\>f_u/Z_A$ & $a\>M_u$ & $a\>f_s/Z_{A}$ &
                       $a\>M_s$ & $f_s/f_u$ \\
\hline

 0.09  & 0.0437(32) & 1.579(19)& 0.0474(30) & 1.595(13)&
1.086(18) \\
 0.10  & 0.0486(31) & 1.375(14)& 0.0529(29) & 1.390(11)&
1.087(15) \\
 0.120 & 0.0609(33) & 0.965(10)& 0.0656(30) & 0.983(8) &
1.077(13) \\
 0.135 & 0.0689(29) & 0.636(6) & 0.0741(25) & 0.655(4) &
1.074(12) \\
 0.145 & 0.0711(24) & 0.382(4) & 0.0757(21) & 0.403(3) &
1.065(10) \\
 0.1492& 0.0580(50) & 0.245(4) & 0.0628(42) & 0.272(3) &
1.082(26) \\
\hline
\end{tabular}
\end{center}
\end{table}

\begin{table}[p]
\caption{\protect{\hspace*{14cm}} \label{table_spacing}}
\vspace{1ex}

The values of the inverse lattice spacing from the string
tension~\cite{BaSch}, using $\sqrt{\sigma} = 420 MeV$.
For comparison we also give the values determined from our
$f_{\pi}$ and $m_{\rho}$ data.

\begin{center}
\begin{tabular}{|c|c|c|c|}
\hline
$\beta$ & $a^{-1}_{\sigma}$ & $a^{-1}_{f_{\pi}}$ &
$a^{-1}_{m_{\rho}}$ \\
\hline
 5.74 & 1.118(9)  & 1.35(8) & 1.41(4)  \\
 6.00 & 1.876(19) & 2.36(8) & 2.15(10) \\
 6.26 & 2.775(18) & 3.46(37)  & 2.94(16) \\
\hline
\end{tabular}
\end{center}
\end{table}
\clearpage
\section*{Figure Captions}

\begin{enumerate}
\item \label{figure_local_mass}
Local masses as defined in eq.~\ref{mloc} for the pseudoscalar
meson
are plotted versus the time separation~$t$ in lattice units.
Shown are
the data for $\beta=$5.74, 6.00 and 6.26 for a light quark mass
of about
$2m_s$ and a heavy quark mass corresponding to $\kappa_h$.

\item \label{figure_ratio_sl}
Ratio of smeared and local correlators as defined in
eq.~\ref{ratio}
at $\beta=6.26$ for different heavy quark masses and a light
quark mass
of about $2m_s$. Deviations from unity signal contamination of
local
correlators by excited states. The scale has been set by use of
$f_\pi$,
cf. table \ref{table_spacing}.

\item \label{figure_ratio_with_beta}
Ratio of smeared and local correlators as defined in
eq.~\ref{ratio}
for $\beta=6.0$ and 6.26 and a heavy quark mass
of about the charm quark mass.

\item \label{figure_fp_extra_a_sigma}
$f_P$ vs $a$ in physical units for different meson masses
fixed by interpolating between the calculated ones. The lines are
linear fits to the data points using only the values where the
ratio
{\KroMac} norm to standard norm is smaller than 1.6.
The scale is set by the string tension.

\item \label{figure_fp_v_mp_inv}
$\hat{f}_P$ is shown as a function of $1/M_P$ at
$\beta=$5.74, 6.00 and 6.26, where the light quark mass is
 extrapolated to the chiral limit. The set of data
points labeled by ``{\KroMac}'' were obtained with the
 normalization eq.~\ref{kronfeld} and the dashed lines denote
the error band after taking the continuum limit of {\it these}
data.
The others denoted by ``Standard" are with the relativistic
normalization. The static points are taken from ref.~\cite{fb3}.
The scale was set
by $\sqrt{\sigma}$.

\item \label{figure_fp_volume}
$f_P/\sqrt{\sigma}$ is shown vs $L\sqrt{\sigma}$ at $\beta=5.74$
for different
heavy quark masses. The light quark mass was fixed at about
$2m_s$.

\item \label {figure_fpi_sigma}
Compilation $f_{\pi}/\sqrt{\sigma}$ as function of $\sqrt{\sigma}
a$. The string tension values are as listed in ref.~\cite{fb3} with
linear interpolation in $\ln (a^2\sigma)$.
The symbols refer to the results of several groups: \\
Filled circle: refs.~\cite{GF11,BLSnew};
Filled square: this work;
Filled triangle: ref.~\cite{GF11};
Filled star: refs.~\cite{BLSnew,Ape} and this work;
Open circle: ref.~\cite{GF11};
Open square: ref.~\cite{UKQCD_light};
Open triangle: this work;
Open rhomb: ref.~\cite{BLSnew};
The star refers to the linearly extrapolated value.

\item \label{figure_fp_extra_mh_fpi}
$\hat{f}_P$ as function $1/M_P$. The scale is taken from $f_{\pi}$.
The lines correspond to the error band described in the text.

\end{enumerate}
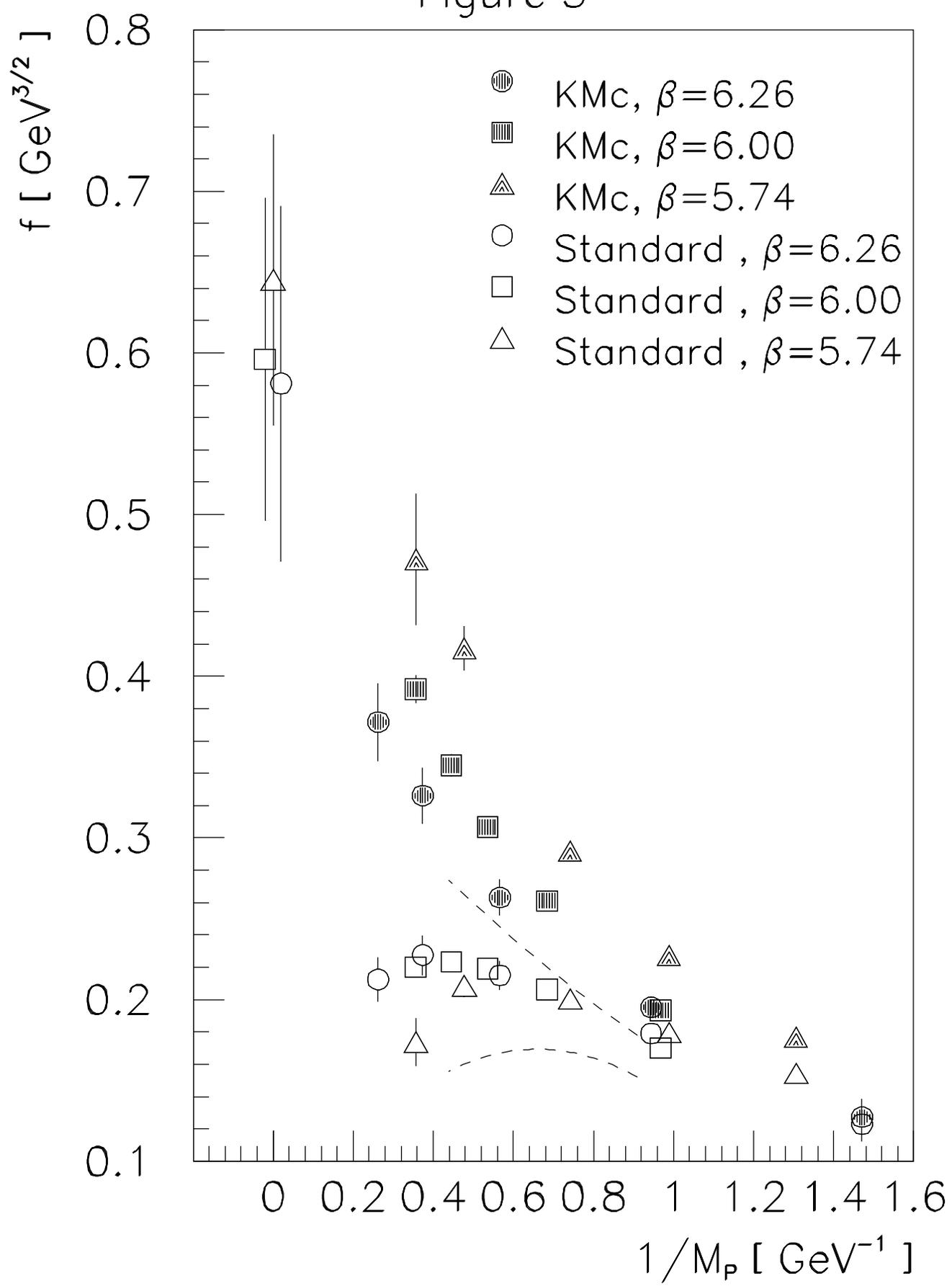
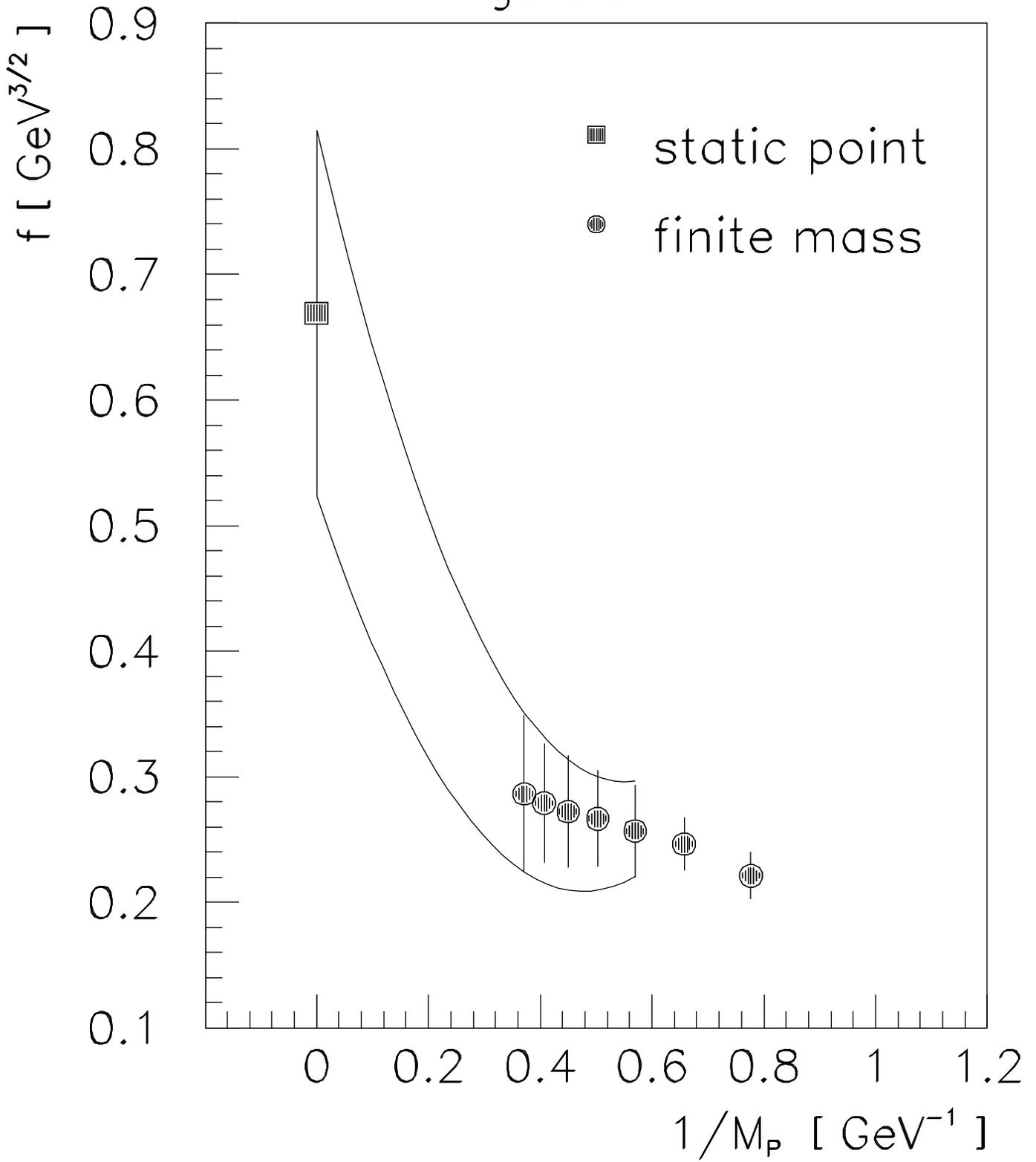
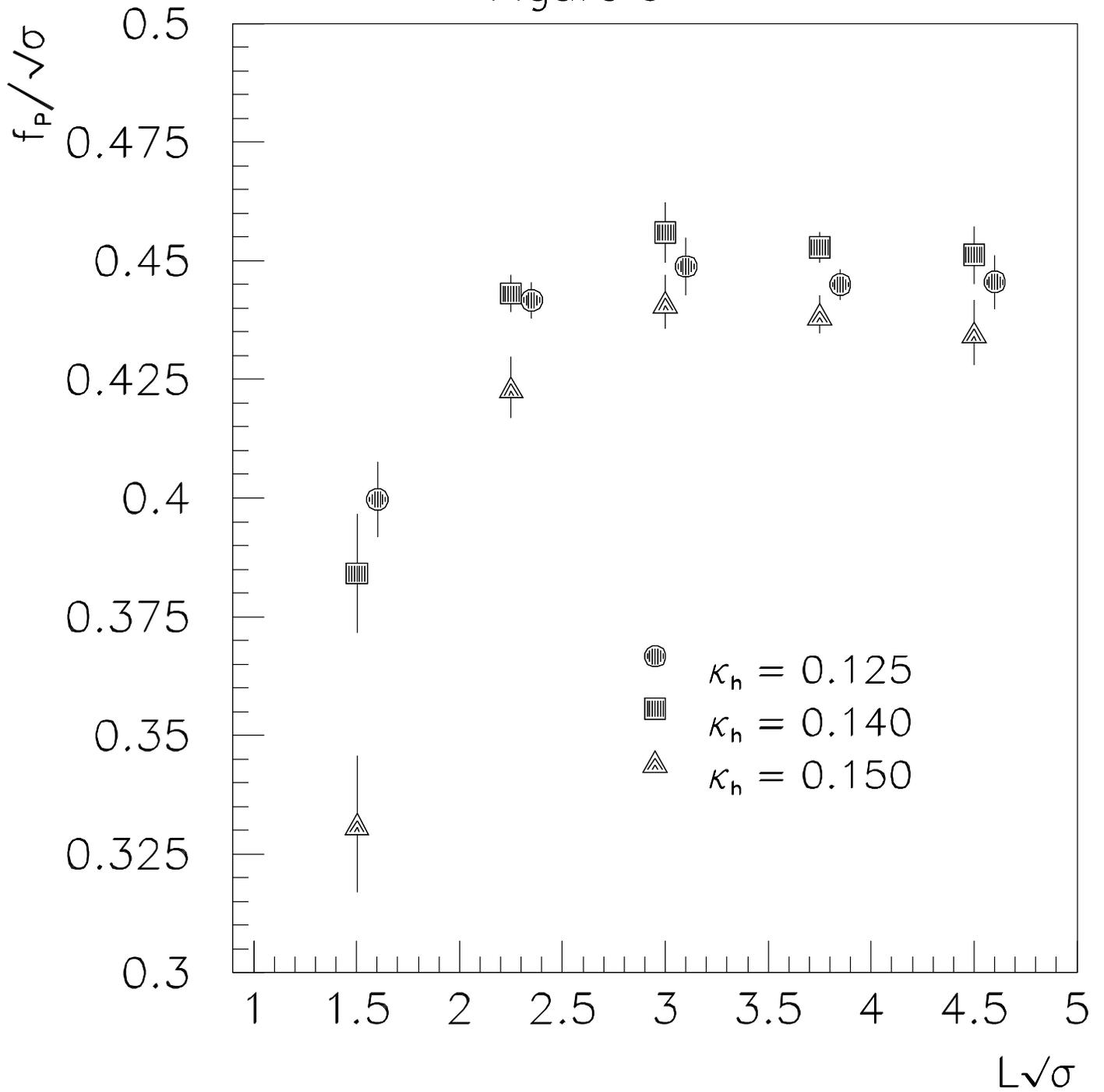
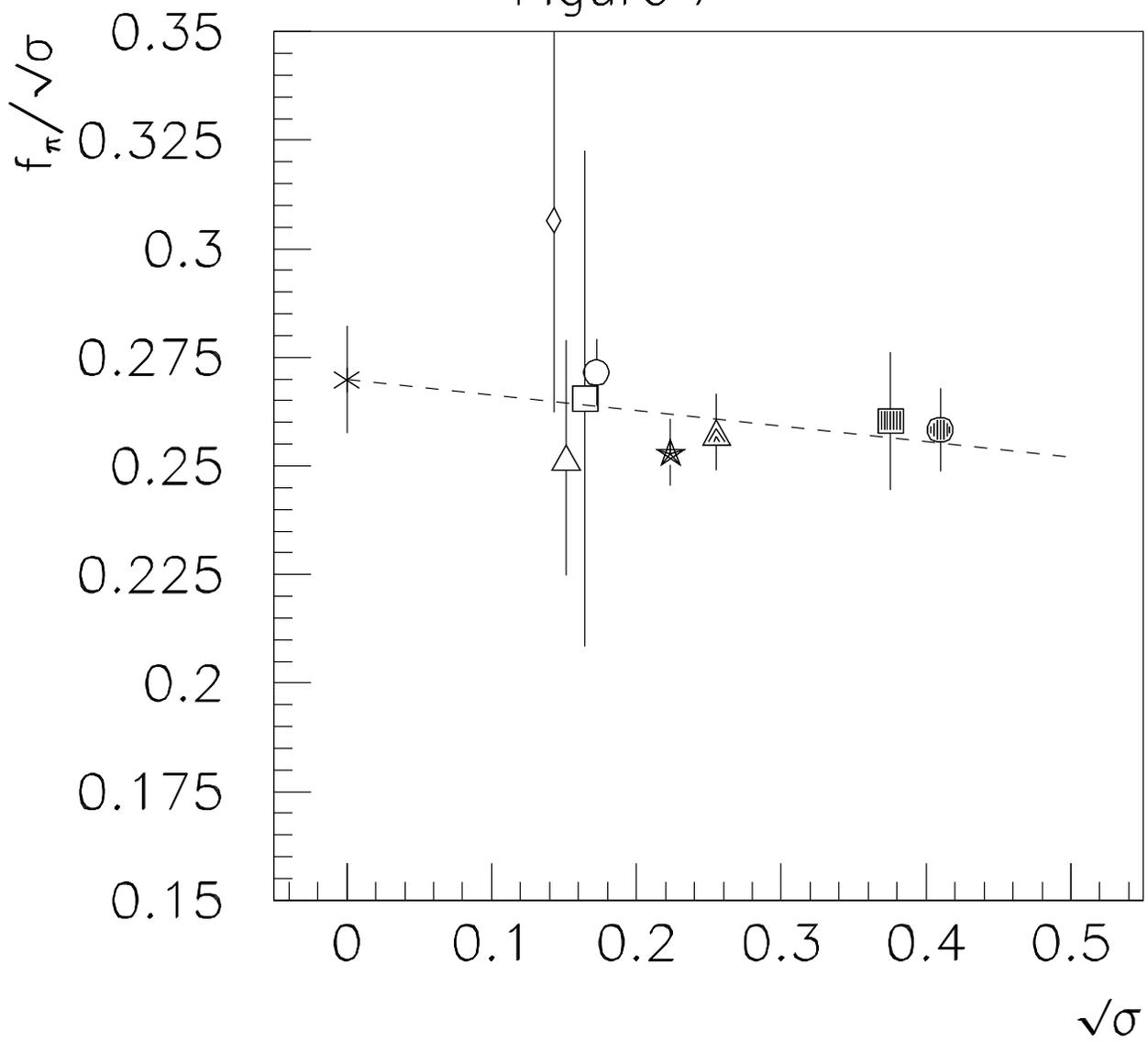
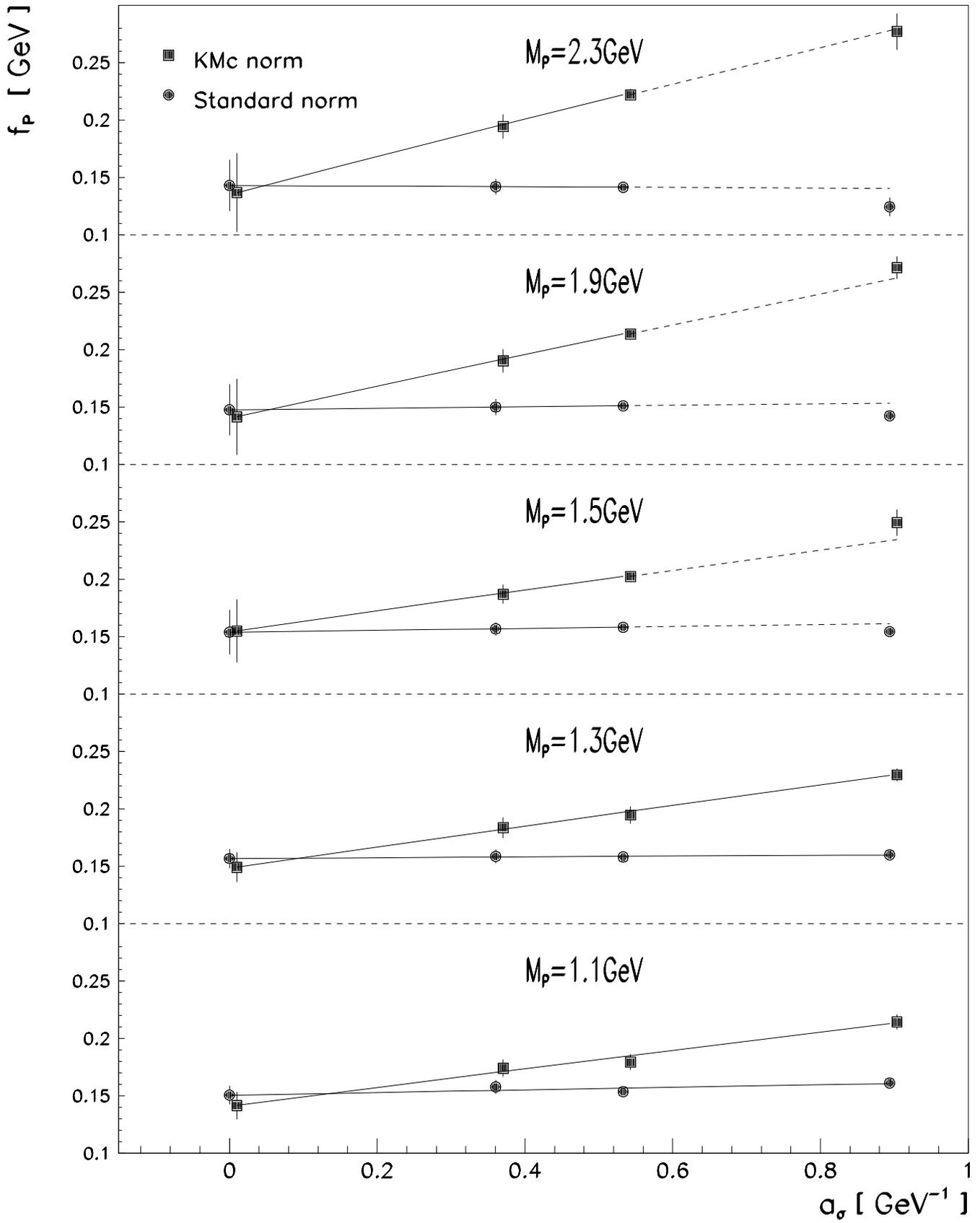
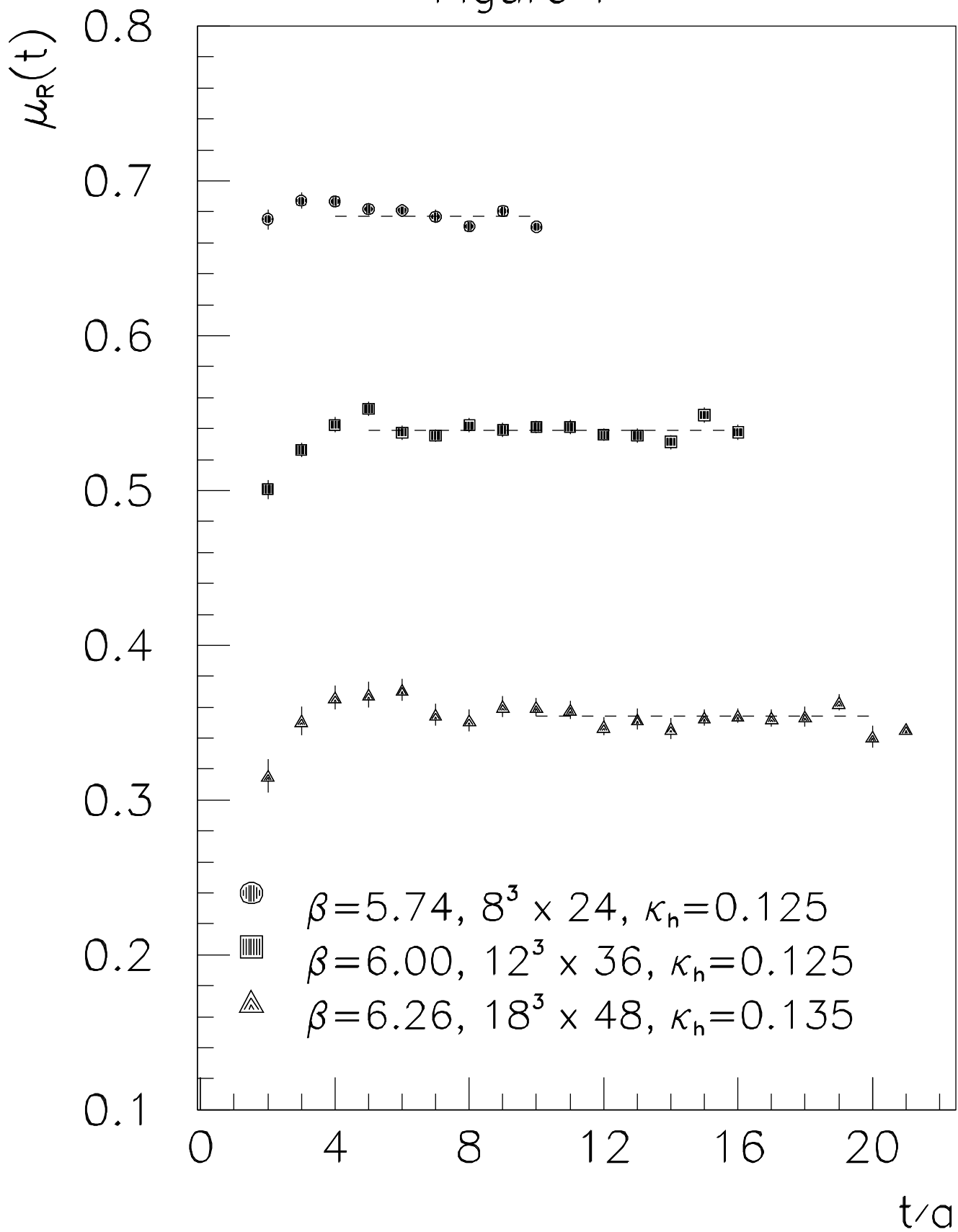
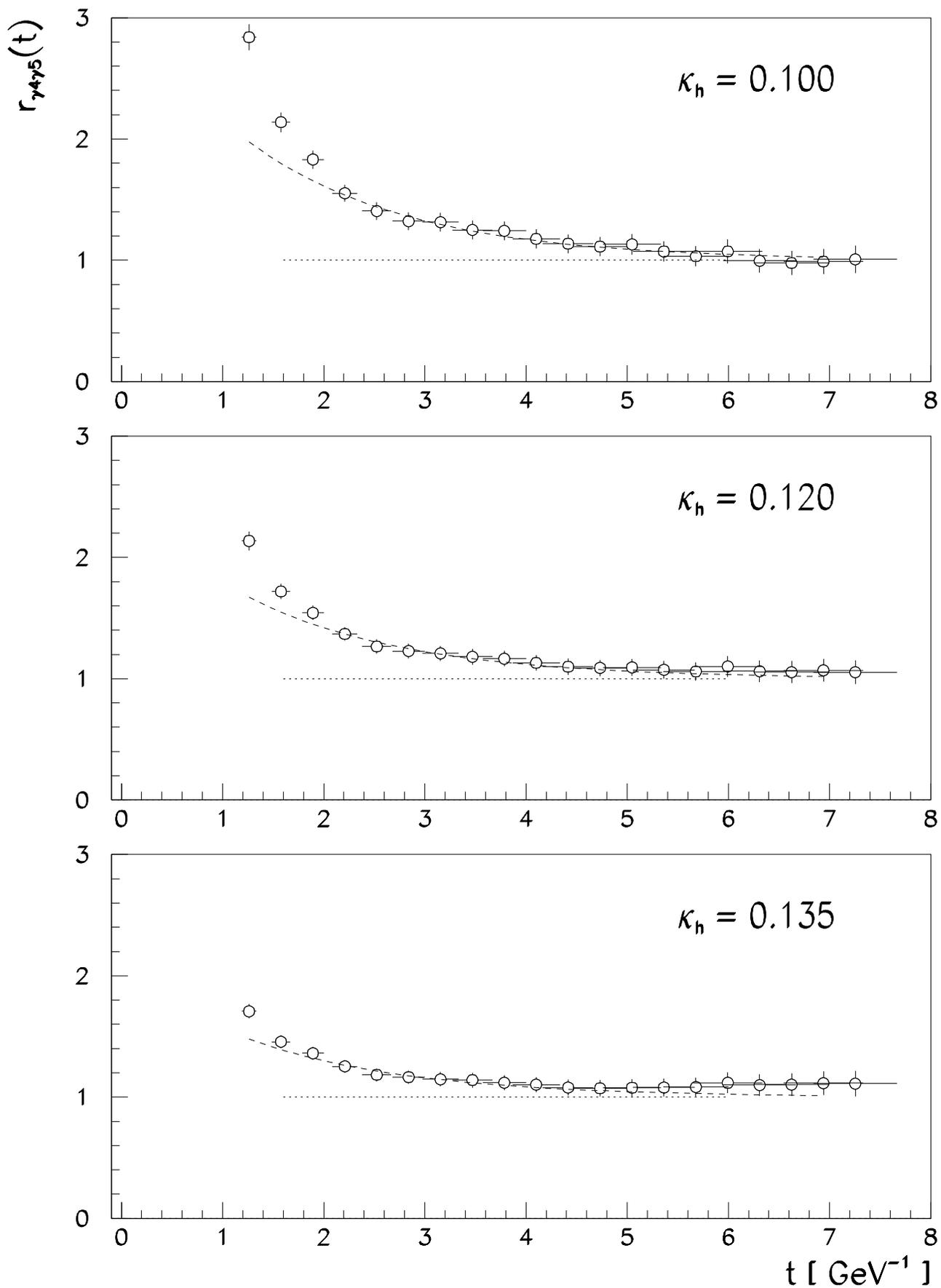
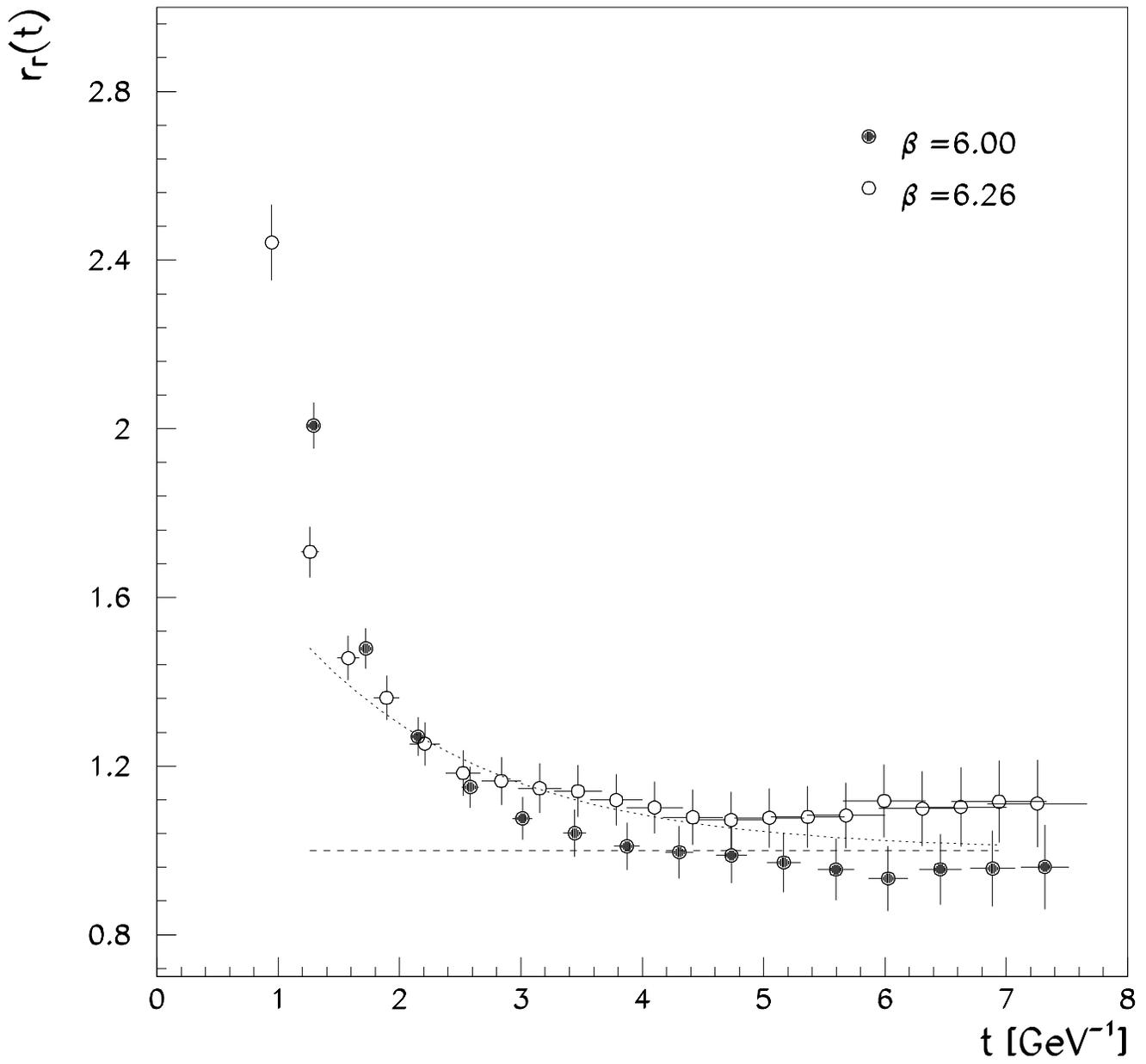
\end{document}